\documentclass[aps,pra,twocolumn,groupedaddress,floatfix]{revtex4-1}

\usepackage{amsmath}
\usepackage{bm}
\usepackage{graphics}
\usepackage{times}
\usepackage{graphicx}
\usepackage{xcolor}
\usepackage{array,multirow,graphicx}
\usepackage{multirow,hhline,graphicx,array}
\usepackage{rotating}
\usepackage{multirow}
\usepackage{booktabs}
\usepackage{hyperref}
\usepackage[markup=underlined]{changes}

\usepackage{graphics}
\usepackage{bm}
\usepackage{color}
\newcommand\Tstrut{\rule{0pt}{2.3ex}}         

\newcommand{\ATTOMESA}{{ATTOMESA}}
\begin{document}
\title{High-Order Harmonic Generation in Helium: A Comparison Study}
\author{A.~T.~Bondy$^{1,2}$}
\email{bondy11u@uwindsor.ca}
\author{S.~Saha$^{2,3}$}
\author{J.~C.~del~Valle$^{2}$}
\author{A.~Harth$^{4}$} 
\author{N.~Douguet$^{5}$}
\author{K.~R.~Hamilton$^{6}$}
\author{K.~Bartschat$^{2}$}
\affiliation{$^1$Department of Physics, University of Windsor, Windsor, ON N9B 3P4, Canada}
\affiliation{$^2$Department of Physics and Astronomy, Drake University, Des Moines, IA 50311, USA}
\affiliation{$^3$Department of Physics, Panskura Banamali College (Autonomous), Panskura, West Bengal 721152, India}
\affiliation{$^4$Center for Optical Technologies, Aalen University, D-73430 Aalen, Germany}
\affiliation{$^5$Department of Physics, University of Central Florida, Orlando, FL 32816, USA }
\affiliation{$^6$Department of Physics, University of Colorado Denver, Denver, CO 80204, USA}


\begin{abstract}
We report a detailed study of high-order harmonic generation (HHG) in helium. 
When comparing predictions from a single-active-electron model with those from all-electron simulations, such as \ATTOMESA\ and $R$-matrix with time-dependence, which can include different numbers of states in the close-coupling expansion, it seems imperative to generate absolute numbers for the HHG spectrum in a well-defined framework. 
While qualitative agreement in the overall frequency dependence of the spectrum,
including the cut-off frequency predicted by a semi-classical 
model, can be achieved by many models in arbitrary units, only absolute numbers
can be used for benchmark comparisons between different approaches.
\end{abstract}

\maketitle
\section{Introduction}
\label{intro}
High-order harmonic generation (HHG) is an important method to 
generate coherent soft X-rays by producing odd harmonics of a fundamental frequency from an intense few-pulse laser, often 
with a frequency in the mid-infrared range \hbox{\cite{hhg_mid_ir_2014,hhg_mid_ir_2016,hhg_mid_ir_2018}}. 
Over the past four decades, numerous experimental and theoretical papers
have been published on the subject -- too many to produce a representative and unbiased 
reference selection of even moderate size. Most studies fall into two categories: the generation of attosecond pulses using HHG~\cite{as_pulse_review}, and the use of HHG as a measurement 
\hbox{tool~\cite{attosecond_spectroscopy_review, corkum_review}}. The most common means of understanding the HHG process is based on the semi-classical ``three-step model'' introduced by Schafer {\it et al.}~\cite{PhysRevLett.70.1599} and Corkum~\cite{corkum1993}, which explains some of the basic
characteristics of the process and its observations.
In that model, after an electron escapes from the target by tunnel ionization, 
it is driven further away by the strong electric field, until it 
is finally accelerated back towards its parent ion upon reversal of the field.  During the recollision process, 
some of the electron's energy is released as a high-energy photon whose frequency must be an odd multiple of
the fundamental frequency of the driving field. The process repeats for various cycles of the driving laser field. Due to the exponential
dependence of the tunnel probability on the width of the barrier, which in turn is determined by the field strength,
the largest effects occur near the peak amplitude of the field.

From the above model, one can estimate the maximum photon energy, or cutoff energy, 
that is producible by HHG from a single laser source. This cutoff energy is given by~\cite{corkum1993} 
\begin{equation}\label{cutoff_eq}
    E_{\rm c} = I_p + 3.17\,U_p,
\end{equation}
where $I_p$ is the atomic ionization potential and $U_p$ is the ponderomotive potential. 
The latter can be approximated in terms of the laser peak intensity and wavelength as
\begin{equation}\label{Up}
    U_p = \frac{2 \, e^2}{c\, \epsilon_0 \,m} \; \frac{I_0 \, \lambda^2}{16 \,\pi^2}.
\end{equation}
Here $e$ is the electron charge, $c$ the speed of light, $\epsilon_0$ the vacuum permittivity, $m$ the electron mass, $I_0$ the laser peak intensity, and $\lambda$ the (central) wavelength of the driving laser. 

Interestingly, the vast majority of papers on the topic of HHG present the spectrum in ``arbitrary units''.
While this might be the only option for experimental studies due to the notorious difficulties associated
with absolute intensity measurements, all the relevant quantities are well defined. Even though different definitions (see below) are used by individual groups,  presenting theoretical HHG spectra with an absolute unit is certainly possible. The quantity generally agreed upon as being the
determining factor for the HHG spectrum is the dipole moment that is induced in the system by the
external driving field.  In particular, we are interested in the Fourier transform of the dipole acceleration, i.e.,
\begin{equation}\label{eq:fourier}
\tilde{a}(\omega) = \int_{-\infty}^{+\infty} a(t) \,{\rm e}^{i \omega t} dt,
\end{equation}
where $a(t)$ is the second derivative with respect to time of the induced dipole moment~$d(t)$.

The present work is a follow-up on our recent study~\cite{Finger2022} of HHG in neon, where we
looked at the particularly challenging problem of reaching the so-called ``water window'' of
HHG frequencies ranging from the $K$-absorption edge of carbon to the $K$-edge of oxygen.
In order to do so, we compared results from a single-active-electron (SAE) model with predictions
from the $R$-matrix with time-dependence (RMT)~\cite{BROWN2020107062} approach.  The latter is
implemented in a general
all-electron code based on the close-coupling formalism.  As such, electron exchange and 
correlation effects are accounted for in RMT, whereas they are neglected in SAE.

While there was qualitative agreement between the SAE and RMT results, it was difficult to
explain the remaining quantitative differences, which were only found due to the fact that
we chose to compare {\it absolute\/} numbers using the formula 
\begin{equation}
S(\omega) \!=\! \frac{2}{3\pi c^3} \,|\tilde{a}(\omega)|^2
\label{eq:joachain}
\end{equation} 
for the so-called ``spectral density''~$S(\omega)$~\cite{Joachain,PhysRevA.87.053406}.
We will refer to the above equation as the ``acceleration form''.
We also investigated equivalent formulas using
the dipole moment directly (the ``length form'') or its first time derivative (the ``velocity form'').
In addition to the order of the time differentiations,
the formulas differ by factors of $\omega^2$ and $\omega^4$ arising from  transforming to the velocity and length forms, respectively.
As a numerical check, we verified that the results from the three methods agree well with one another, although the spectra are not exactly equivalent due to additional terms that may be present after the pulse has vanished (see Eq.~(22) of~\cite{Guan2006}). All of the results in this paper are presented in the acceleration form, which is the generally preferred form, in particular for short pulses~\cite{PhysRevA.79.023403,PhysRevA.81.063430}.
We emphasize again that the pre-factor in the above equation, 
whose numerical value is approximately $8.25 \times 10^{-8}$ in atomic units 
together with the definition in Eq.~(\ref{eq:fourier}),
provides an {\it absolute\/} value for the quantity of interest. Unless indicated otherwise, 
atomic units (\hbox{$\hbar \! = \! e = \! m \! =\!1$}; \hbox{$c\!\approx\!137$)}, are used throughout this manuscript.

As mentioned above, other authors use different definitions. Specifically, we note the papers by
Tong and Chu~\cite{Tong2001} and Guan {\it et al.}~\cite{Guan2006} who defined their spectrum by
\begin{equation}
P(\omega) \!=\! \frac {1}{\omega^4 \tau^2} \,|\tilde{a}(\omega)|^2,
\label{eq:guan}
\end{equation} 
where~$\tau$ is the length of the pulse.  The authors used a fixed number of cycles 
with a $\sin^2$ envelope for the electric field.  In this case, 
the pulse length~$\tau$ from the very beginning to the very end is well defined, in contrast to somewhat
more realistic pulses, e.g., a Gaussian envelope, where one usually defines the length via the full-width
at half-maximum of the intensity.  Normalizing to the pulse length has the advantage that pulses with a different
number of cycles can be better compared with respect to the yield.  Note, however, that the frequency dependence
is very different in the two definitions, since the factor~$1/\omega^4$ now appears in the acceleration form
of Refs.~\cite{Guan2006,Tong2001}. 

In the calculations reported in this paper, we also 
used a linearly polarized laser pulse  with a $\sin^2$-envelope, specifically
\begin{equation}
\boldsymbol{E}(t)\ =\ 	\sqrt{\frac{2I_0}{\epsilon_0\,c}}\sin^2\left(\frac{\omega t}{2N} \right)\cos(\omega t)\hat{\boldsymbol{z}}\ .
\label{laser}
\end{equation}	
The electric field is assumed to vanish outside of the interval $0 < t < 2\pi N/\omega$. We used timesteps between 0.006 and 0.02 in the various methods described below to ensure that the propagation yielded a sufficiently accurate value of the induced dipole moment, which is the fundamental quantity of interest.  

To simplify matters in calculating absolute HHG spectra, we decided to perform a comparison study using helium rather than the much
more complex neon target. In this case, one might expect the SAE and RMT results to agree reasonably well, especially
if the RMT model only includes a single target state, namely He$^+(1s)$.  This corresponds effectively to
the ``static exchange'' approximation for electron scattering from He$^+$. In the static exchange model, we also added results from a new multi-electron time-dependent code, \hbox{\ATTOMESA}, to give us an additional source of comparison.
We emphasize that our calculations, like many other theoretical attempts, are single-atom simulations. While they are important in practical realizations of the HHG process, we are not concerned with macroscopic effects such as phase matching or volume averaging of the intensity in the present work.  We note, however, that work in this direction using the RMT code was recently reported~\cite{Hutcheson_2023}.

Furthermore, it is possible to 
systematically test the effects of extending the close-coupling  model by including more states, such as the
$n=2$ and $n=3$ states of He$^+$. Such models should indicate the importance of channel coupling as well as 
correlations in the initial state, whose expansion then includes doubly-excited states.  

This manuscript is organized as follows.  In section~\ref{sec:theory}, we briefly describe the SAE,
RMT, and \hbox{\ATTOMESA} models used in the present work.  This is followed by the presentation and discussion 
of our results for two wavelengths,
248.6$\,$nm and 1,064$\,$nm.  Since these are the wavelengths investigated by Tong and Chu~\cite{Tong2001}
and Guan {\it et al.}~\cite{Guan2006}, we selected them and the intensities chosen in the latter studies to have a further basis for comparison with other works. As will become clear below, some of these laser parameters appear to be quite challenging from a numerical perspective.  Consequently, we judge them to be a good start for benchmark studies, even though their Keldysh parameters are not in the predominantly tunneling regime.

As will be shown below, the {\it quantitative\/} calculation of HHG spectra is
by no means trivial, due to the well-known fact that the conversion efficiency of the process is often very
low.  Most of the emitted radiation may simply come back at the fundamental
frequency, although the choice of laser parameters (if available in the setup) can change this.
Even though 248.6$\,$nm is not a particularly suitable wavelength
for HHG in practice, the calculation in Ref.~\cite{Guan2006} was performed with a sophisticated two-electron 
code specifically designed for the helium target. 
These benchmark results, fortunately also given as absolute numbers, provide an excellent further source of comparison.  
Such a  thorough test is highly desirable to assess the reliability of all the results presented here, as well as those from other approaches, for longer wavelengths and higher intensities. 

\section{Theory}
\label{sec:theory}
Since the SAE and RMT methods have been described in previous papers, we 
limit ourselves to a brief summary and the specifics of their applications 
to the helium target. We also add a short description of the \hbox{\ATTOMESA} code, which is currently under development. 

\subsection{The SAE model}
\label{subsec:SAE}
We employed the same SAE model as Birk {\it et al.}~\cite{Birk2020} and Meister {\it et al.}~\cite{PhysRevA.102.062809}. 
Specifically, we used the one-electron potential 
\begin{equation}\label{eq:potential}
V(r) = -\frac{1}{r} - \left(\frac{1}{r} + 1.3313\right) \,\exp(-3.0634\,r),
\end{equation}
where $r$ is the distance from the nucleus, to calculate the valence orbitals. 
We used a variable radial grid with a smallest stepsize of 0.1 near the origin, gradually increasing by 5\% per step up to a maximum stepsize of 0.2 to a box size
of approximately 200.  
In this discretization, the difference between the predicted
ionization potential and the excitation energies of all 
(1s$n\ell$) states with singlet spin character (the SAE is, of course, only concerned with the $n\ell$ orbital) up to $n\!=\!6$ and $\ell \!=\! 5$ 
with the recommended values from the NIST database~\cite{NIST} is less than
0.2$\,$eV even in the worst-case scenario.

Assuming that the driving radiation is linearly polarized,  
the initial state can be propagated very efficiently and accurately. We used an updated 
version of the code described by Douguet
{\it et al.}~\cite{PhysRevA.93.033402}. Based on many previous works using SAE approaches, we expected it to 
be more suitable for the helium target compared to heavier noble gases, as long as obvious two-electron correlation effects, 
e.g., auto\-ionizing resonances, are not affecting the process significantly. 
Most of the calculations were performed with partial waves up to a total angular momentum of~47. Spot checks carried out with angular momenta up to 120 did not change the results reported below significantly. 

We then calculated the induced dipole moment as 
\begin{equation}\label{eq:dipole}
d(t) = \langle z \rangle(t) = \langle \Psi(\bm{r},t) | z | \Psi(\bm{r},t)\rangle 
\end{equation}
during the pulse and numerically differentiated the function twice to obtain the dipole acceleration required in Eq.~\ref{eq:fourier}. Since there are two electrons
in the $1s$ orbital of helium, the above dipole moment should be multiplied by~2 to account for the occupation number~\cite{Tong2001}.

\subsection{RMT}
\label{subsec:RMT}
As a second method, we employed the general \hbox{$R$-matrix} with time-dependence (RMT)
method~\cite{BROWN2020107062}. RMT has been applied to the study of HHG several times, including HHG from two-color fields~\cite{hamilton_two_colour}, XUV-initiated HHG~\cite{brown_xuvhhg}, and HHG from mid-IR lasers~\cite{ola_nearIR, Finger2022}.  Most recently, it has been extended to include macroscopic propagation effects in HHG~\cite{Hutcheson_2023}. 

To calculate the necessary time-independent basis functions and dipole matrix elements for the present work, 
we started with the simplest possible model, namely a non\-relativistic \hbox{1-state} 
approach. This model, labeled \hbox{RMT-1st} below, is the one most closely related to the SAE approach.  
It essentially describes electron collisions with He$^+$ in the ionic ground state.
Nevertheless,
there are some subtle differences, for example in the $1s$ orbital used in the two models. In the SAE, the potential given
in Eq.~(\ref{eq:potential}) 
supports bound states, and the $1s$ orbital obtained that way is close to the Hartree-Fock orbital of the ground-state configuration.  
Since the potential used in the SAE calculations, given in Eq.~(\ref{eq:potential}), is not {\it ab initio}, the resulting ground-state energy of $-2.910$ is slightly lower than obtained in
non\-relativistic variational calculations~\cite{Drake-Helium}.
In RMT, on the other hand, we use the analytically known $1s$ orbital of He$^+$. While this is not optimal 
to obtain the best ground-state energy, it is a suitable orbital to describe essentially all excited states of neutral helium,
which are automatically contained (to a very high accuracy) in the close-coupling approach, as long as the state fits into the
\hbox{$R$-matrix} box.

In addition to \hbox{RMT-1st}, we set up two more models, one where we added the $n\!=\!2$ states of He$^+$ using the 
$2s$ and $2p$ ionic orbitals and another one where we also included the $n\!=\!3$ states.  These two models will be referred to as 
\hbox{RMT-3st} and \hbox{RMT-6st}, respectively. We used an \hbox{$R$-matrix} radius of 40.0 and 60 $B$-splines to expand
the continuum orbitals inside the $R$-matrix box. Partial waves up to a total orbital angular momentum of 49 were used to ensure
converged results for all cases reported below. Since we noticed a few surprising results, to be discussed in the next section, we varied the size of the \hbox{$R$-matrix} box, the number of splines, the number of partial waves, as well as the radial grid and the time step.
We even altered the numerical accuracy of the real variables in the computer code. The results reported below did not change significantly.  
They were obtained with 60 splines employed to expand the one-electron orbitals inside the $R$-matrix radius of 40 and an outer-region size of up to 8,000. The results did not change significantly whether or not an absorbing boundary condition was applied near the outer edge. 
  
\subsection{ATTOMESA}
\label{subsec:ATTOMESA}

As a third approach, we utilize \ATTOMESA, a novel time-dependent multi-electron code tailored for applications in attosecond science, particularly focusing on the treatment of atoms and molecules under intense fields. This methodology bears resemblances to the time-dependent restricted-active-space configuration-interaction method \cite{Hochstuhl2012}, \hbox{XCHEM~\cite{Marante2014,Marante2017}}, and ASTRA~\cite{Randazzo2024} numerical techniques. In this study, we present the initial outcomes of \ATTOMESA~and will concentrate our discussion on its most salient features. A comprehensive description of the method will be provided in a forthcoming communication.

In \ATTOMESA, short-range electron correlations and exchanges are described from a set of Gaussian-type orbitals (GTOs) centered on each atom. The latter space is augmented with finite-element discrete-variable (FEDVR) functions~\cite{Rescigno2000}, in contrast  to the \hbox{B-Spline} functions employed in XCHEM, ASTRA, and RMT. These FEDVR functions are subsequently orthogonalized to all GTOs, thus establishing a mixed GTO-FEDVR basis set that enables an accurate description of short-range multi-electron interactions and the electronic continua associated with diverse excited electronic states of the residual ion. 

In contrast to XCHEM, which employs a single-center expansion for computing electronic integrals involving \hbox{B-Spline} functions and GTOs, our approach utilizes the Becke integration scheme~\cite{Becke1988,Becke-Heman} for evaluating electronic integrals involving FEDVRs and GTOs within the inner region defined by the spatial extension of the GTOs. Beyond this region, where the photo\-electron dynamics is described solely by FEDVRs, the electronic integrals are computed in a single-center grid using Gauss-Lobatto quadrature~\cite{Rescigno2000}. This method capitalizes on the unique properties of FEDVRs, particularly their straightforward representation of any local operator. Consequently, it results in an exceptionally sparse atom-field interaction Hamiltonian in the length gauge, facilitating the development of efficient time-propagation schemes. The exponential time propagator is further decomposed into field-free and atom-field interaction operators, with the action of the field-free Hamiltonian handled in its diagonal representation, while the atom-field interaction is expressed in the FEDVR representation.

The results obtained from \ATTOMESA~have been compared against SAE predictions for HHG from atomic hydrogen, utilizing a hybrid Gaussian-FEDVR basis set. Virtually perfect agreement was found between both methods, thereby validating the computation of the electronic integrals and the time-propagation method employed in \ATTOMESA. Furthermore, the non\-relativistic energies of the Rydberg states of atomic hydrogen, extending up to $n=15$, were reproduced with a precision up to ten decimal places. Nevertheless, \hbox{\ATTOMESA} is still under development. The calculation of the integrals will be drastically improved, and parallel-processing capabilities are not yet supported. For this reason, we are still limited in the parameter space where \ATTOMESA\ can be applied.  Therefore, calculations were only performed for the shorter wavelength and lower-intensity pulses considered in this study. The principal goal was to assess the consistency between different methods and models.

In our calculations, we use the helium \hbox{cc-pVQZ} basis set~\cite{cc-pVQZ} and construct our reference space from the first $s$ and $p$ orbitals obtained from a Hartree-Fock calculation for He. This space is complemented with the $1s$ orbital of He$^+$. We obtained the He ground state energy as $-2.900$ and virtually the exact He$^+$ ground-state energy of $-2.0$. We use a box size of 300, with a complex absorbing potential (CAP) placed near the edge, and elements of radial size 10 with ten DVRs per element. For the present work, we only performed static exchange (one-state) calculations with \hbox{\ATTOMESA}.  The largest angular momentum included in the time propagation was $\ell=15$. These parameters are sufficient for the short wavelength of 248.6~nm and intensity of $1 \times \rm 10^{14}\,W/cm^2$ for which we report \hbox{\ATTOMESA} results.

\section{Results and Discussion}
\label{sec:results}
We now discuss our results for two wavelengths.  The first one (248.6~nm) was chosen by
Guan {\it et al.}~\cite{Guan2006} using a sophisticated computer code
specifically designed for the two-electron helium target.  Due to the complexity of the model
and the available computational resources at the time,
the wavelength and pulse duration were chosen relatively short in order to produce benchmark
results. 

The second wavelength (1,064~nm) was chosen by Tong and Chu~\cite{Tong2001}, but with an SAE potential 
generated via density-functional theory. Hence, comparing with their results serves as an indicator
for the sensitivity of the predictions to the potential employed in the SAE calculation.  Most importantly, however,
in light of the rapid developments in computational power, 
we could employ the RMT model also for this more challenging wavelength and longer pulses.

\subsection{248.6$\,$nm Driving Field}
\label{subsec:248p6}

\begin{figure}[b!]
\resizebox{0.480\textwidth}{!}{\includegraphics{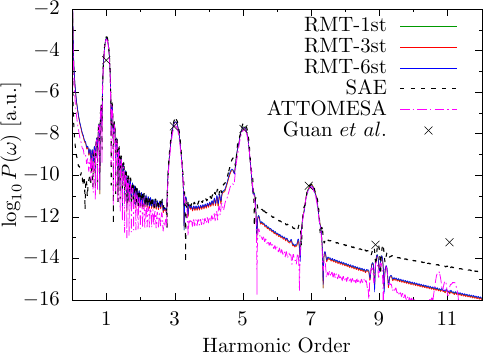}
}
\caption{
Spectrum for a 15-cycle pulse of central wavelength 248.6$\,$nm with a sin$^2$ envelope 
of the electric field and peak intensity of $1 \times \rm 10^{14}\,W/cm^2$.  The results from our SAE model
are compared with predictions from \ATTOMESA\ and three RMT calculations with a different number of coupled states.
Note that the RMT-1st and RMT-3st results are practically indistinguishable.
The predicted cut-off energy from the three-step model is $\approx 26\,$eV, corresponding to the 5$^{\rm th}\,$harmonic. The crosses are the results for the peak values obtained by Guan {\it et al.}~\cite{Guan2006} in the acceleration form.
}
\label{fig:fig1}       
\end{figure}

\begin{figure}[t!]
\resizebox{0.480\textwidth}{!}{\includegraphics{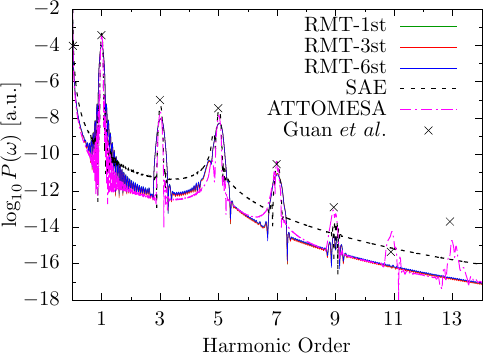}
}
\caption{
Spectrum for a 30-cycle pulse of central wavelength 248.6$\,$nm with a sin$^2$ envelope 
of the electric field and peak intensity of $1 \times\rm 10^{14}\,W/cm^2$.
Again, the RMT-1st and RMT-3st results are practically indistinguishable.
}
\label{fig:fig2}       
\end{figure}

\begin{figure}[t!]
\resizebox{0.480\textwidth}{!}{\includegraphics{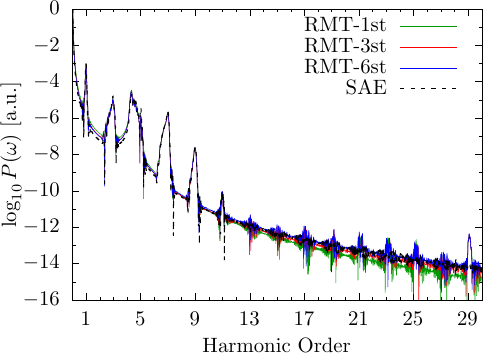}
}
\caption{
Spectrum for a 30-cycle pulse of central wavelength 248.6$\,$nm with a $\sin^2$ envelope 
of the electric field and peak intensity of $5 \times \rm 10^{14}\,W/cm^2$.
The predicted cut-off energy from the three-step model is \hbox{$\approx\,34\,$eV}, corresponding to the 7$^{\rm th}\,$harmonic.
}
\label{fig:fig3}       
\end{figure}

Figures~\ref{fig:fig1} --~\ref{fig:fig3} show the HHG spectrum for 248.6$\,$nm, using the style
of Guan {\it et al.}~\cite{Guan2006}.
Unfortunately, their results are not available in numerical form.  We digitized their data, but an additional curve would make the figure practically unreadable.  However, a close visual inspection shows good agreement in the main features produced by all numerical methods, i.e., the heights of the fundamental peak and of the first few harmonics. The exception is the height of the fundamental peak of Guan {\it et al.} in Fig.~\ref{fig:fig1}, where their result in the acceleration form seems to lie significantly below that obtained in the length form.  The latter agrees well with our predictions (see their Fig.~5). 

Nevertheless, there are several observations worth commenting on. To begin with, the \hbox{RMT-1st} and \hbox{RMT-3st}
results are very similar. However mainly at the minima between the peaks, the results from the other models differ, with the SAE numbers close but certainly not identical to those from \hbox{RMT-1st} and \hbox{RMT-3st}, while the \hbox{RMT-6st} results are significantly different. Upon further analysis, we confirmed that the main reason for the change from \hbox{RMT-3st} to \hbox{RMT-6st} is the inclusion of the He$^+(3p)$ state in the close-coupling expansion. 

We performed exhaustive tests to check the numerical accuracy of the results.
None of the changes made in the numerical treatment altered the numbers for the dipole moments, and subsequently the resulting harmonic spectra, significantly.
Since the dipole moment is the underlying physical quantity that determines the spectrum,
and it apparently converges well against changes in the model (see also the discussion in the next section),
we do believe that the basic physics is contained in our treatment.  The {\it quantitative\/} calculation of 
the HHG spectrum, on the other hand, presents a surprisingly difficult task.

\begin{figure}[t!]
\resizebox{0.480\textwidth}{!}{\includegraphics{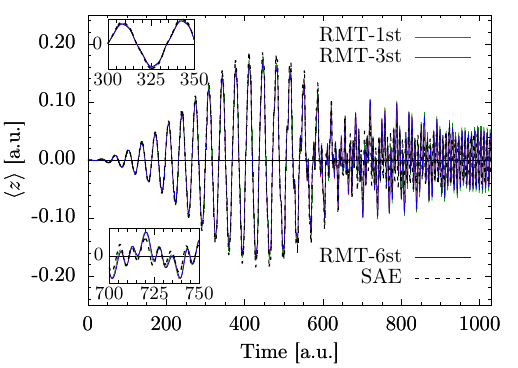}
}
\caption{
Dipole expectation value as a function of time for a 30-cycle pulse of central wavelength 248.6$\,$nm with a sin$^2$ envelope 
of the electric field and peak intensity of $5 \times \rm 10^{14}\,W/cm^2$ (same laser parameters as Fig.~\ref{fig:fig3}).
}
\label{fig:fig3dipole}       
\end{figure}

As explained above, \hbox{\ATTOMESA} cannot yet describe processes for the high-intensity and long pulses used in Fig.~\ref{fig:fig3}. The remaining part of this paper is, therefore, devoted to a further analysis of the various RMT predictions.

In the interest of benchmarking and testing computer codes, it is advisable to look at the quantity that actually 
determines the HHG spectrum, independent of how one might define it.  As an example, Fig.~\ref{fig:fig3dipole}
shows the induced dipole moment for the spectrum exhibited in Fig.~\ref{fig:fig3}.  Once again, there are several interesting 
observations to be made. First of all, it is very difficult to see differences in the dipole expectation value predicted 
by the various models.  This is due to the fact that the fundamental frequency is very dominant, i.e., all the higher harmonics
are due to tiny differences from the dipole moment adiabatically following the driving field.  

\begin{figure}[t!]
\resizebox{0.480\textwidth}{!}{\includegraphics{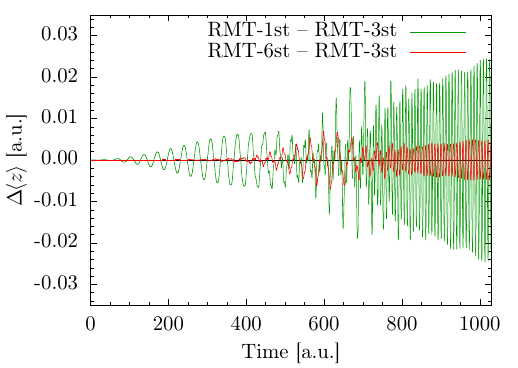}
}
\caption{
Difference between the dipole expectation values obtained with the \hbox{RMT-1st} and \hbox{RMT-6st} models relative to
the results obtained with \hbox{RMT-3st}  as a function of time for the laser parameters used to produce Fig.~\ref{fig:fig3} (30-cycle pulse of central wavelength 248.6$\,$nm with a sin$^2$ envelope 
of the electric field and peak intensity of $5 \times \rm 10^{14}\,W/cm^2$).  
}
\label{fig:fig3dipolediff}       
\end{figure}

We therefore exhibit the {\it differences\/} between the dipole expectation values calculated within the various models in Fig.~\ref{fig:fig3dipolediff}.  As seen there, the maximum deviations in 
the predicted induced dipole moment 
by the \hbox{RMT-1st} and \hbox{RMT-6st} from the \hbox{RMT-3st} numbers  
(which we selected as the reference) are more than an order of magnitude smaller than the values of the dipole moment
themselves.  This already suggests that the latter must be calculated to very high accuracy in order to obtain reliable {\it absolute\/}
values for the HHG spectrum. 

A second feature of interest concerns the development of the dipole moment shortly after the maximum amplitude of the 
driving field.  The regular structure seen up to then suddenly becomes very irregular, and the dipole moment certainly
does not go back to (near) zero at the end of the pulse.  In fact, the oscillations occur with a much higher frequency than 
that provided by the driver.  We analyzed this behavior and noticed that it is due to a remaining small (a few percent)
occupation of the He$\,(1s2p)^1P^o$ state.  The excitation energy
is between 4 and 5 driving photon energies above the ground state.  Nevertheless, at the end of the pulse, the system is essentially in a 
coherent superposition of the ground state and the He$\,(1s2p)^1P^o$ state, with small additions from other excited 
bound states.  As a result, we see oscillations with a frequency corresponding to the energy difference between these two states.
A close inspection of the spectra exhibited in Figs.~\ref{fig:fig1} --~\ref{fig:fig3} reveals that these oscillations indeed
affect the appearance of the spectra near the 5$^{\rm th}$ harmonic, where some additional structures can be seen.

In our previous paper~\cite{Finger2022}, we discussed the effect of using windows to numerically force the induced dipole 
moment to zero at the end of the pulse.  The above discussion shows that this will not only affect, once again, the
{\it absolute\/} values and the details of the spectrum, but it is in fact unphysical. 

\subsection{1,064$\,$nm Driving Field}
\label{subsec:1064}

\begin{figure}
\resizebox{0.480\textwidth}{!}{\includegraphics{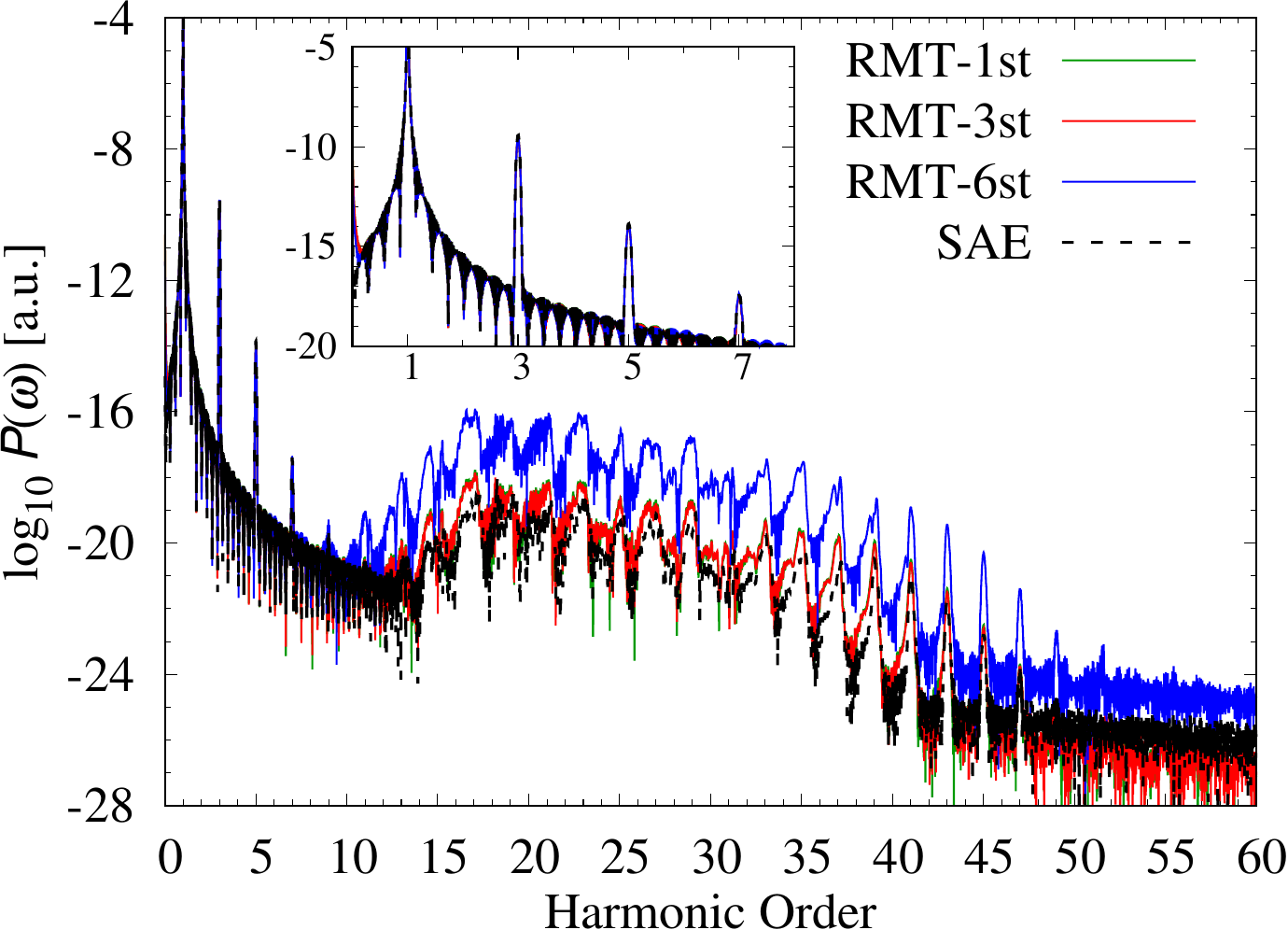}}
\caption{
Spectrum for a 60-cycle pulse of central wavelength 1,064$\,$nm with a sin$^2$ envelope 
of the electric field and peak intensity of $\rm 5 \times 10^{13}\,W/cm^2$.  The results from the SAE model
are compared with predictions from three RMT calculations with a different number of coupled states.
The predicted cut-off energy from the three-step model is \hbox{$\approx\,41\,$eV}, corresponding to the 35$^{\rm th}\,$harmonic.
}
\label{fig:fig4}       
\end{figure}

\begin{figure}[t!]
\resizebox{0.480\textwidth}{!}
{\includegraphics{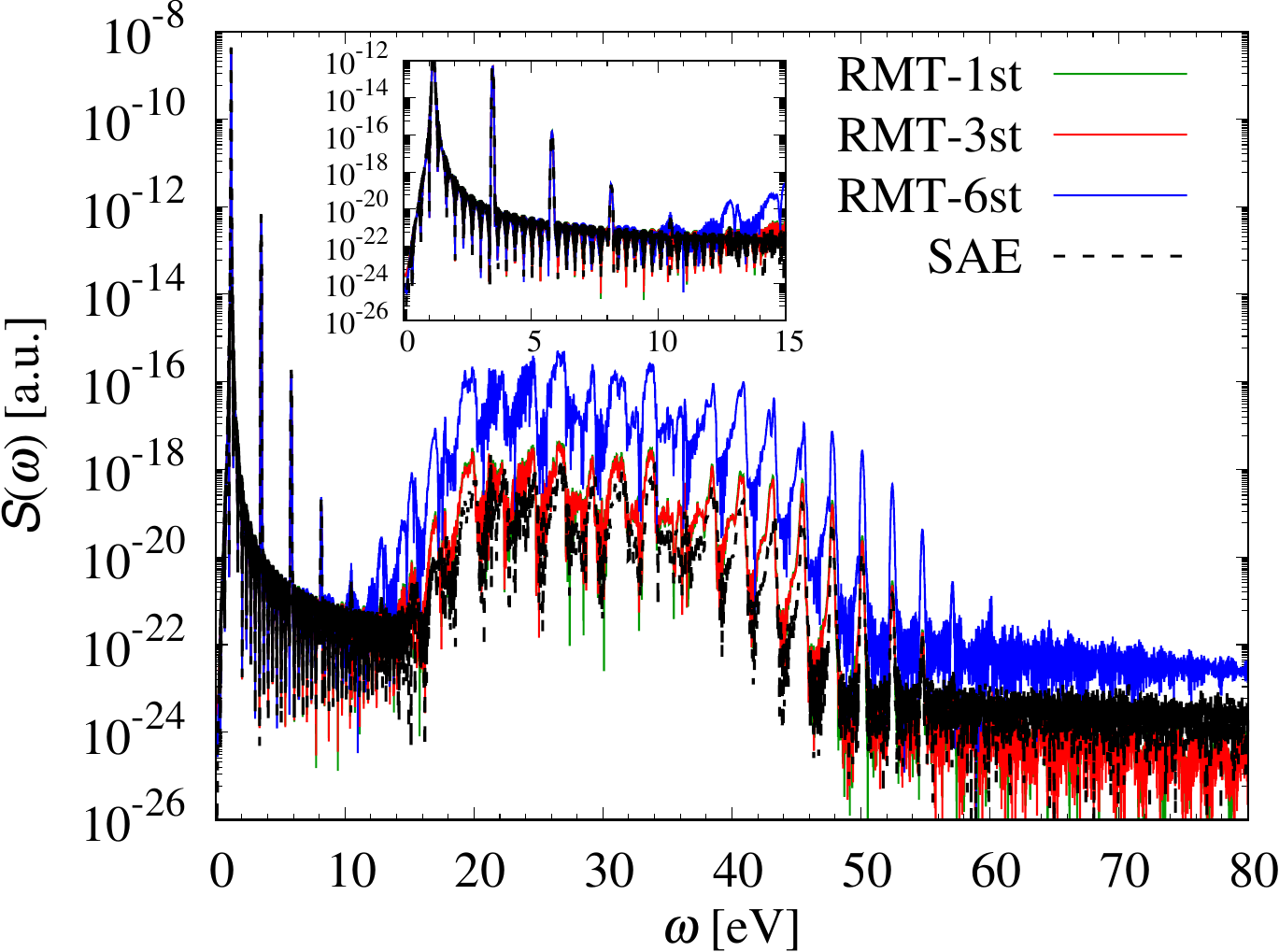}}
\caption{
Spectral density as defined in Refs.~\cite{Joachain,PhysRevA.87.053406} for a 60-cycle pulse of central wavelength 1,064$\,$nm with a sin$^2$ envelope of the electric field and peak intensity of $\rm 5 \times 10^{13}\,W/cm^2$ (the parameters of Fig.~\ref{fig:fig4}). The $y$-axis is logarithmic.
}
\label{fig:fig4b}       
\end{figure}

Figures~\ref{fig:fig4} and~\ref{fig:fig4b} show the HHG spectrum for a 60-cycle pulse with a central
wavelength of 1,064$\,$nm and  
a sin$^2$ envelope of the electric field with a peak intensity of $\rm 5 \times 10^{13}\,W/cm^2$.  
This is one of the cases reported by Tong and Chu~\cite{Tong2001}.  In contrast to the spectra shown in the previous section, which diverge at very small frequencies due to the $1/\omega^4$ in their
definition, the ones exhibited in
Fig.~\ref{fig:fig4} appear to remain finite on the grid shown. Nevertheless, we also show the spectral density in Fig.~\ref{fig:fig4b} after evaluating 
it according to Eq.~(\ref{eq:joachain}).  

Once again, the \hbox{RMT-6st} results are significantly different than those obtained in the other models. 
In particular, the height of the plateau is significantly different in the RMT-6st calculation compared to the other models. The fraction of the harmonic spectra under the first few odd harmonics and the aforementioned plateau is summarized in Table~\ref{tab:tab1} below.

\begin{figure}[t!]
\resizebox{0.480\textwidth}{!}{\includegraphics{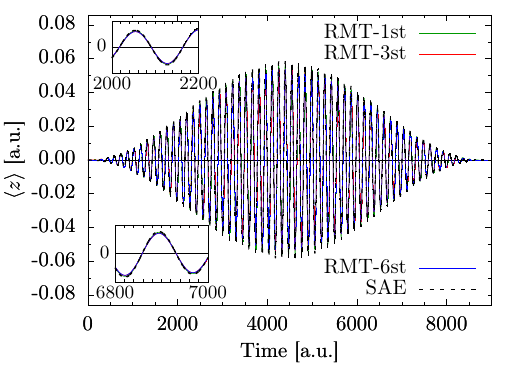}
}
\caption{
Dipole expectation value as a function of time for a 60-cycle pulse of central wavelength 1,064$\,$nm with a sin$^2$ envelope of the electric field and peak intensity of $\rm 5 \times 10^{13}\,W/cm^2$ (the parameters used to produce Fig.~\ref{fig:fig4}).
}
\label{fig:fig4dipole}       
\end{figure}

\begin{figure}[t!]
\resizebox{0.480\textwidth}{!}{\includegraphics{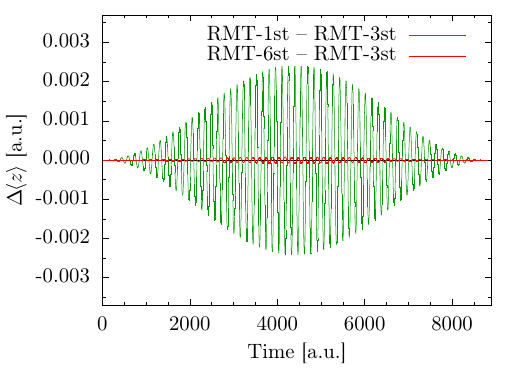}
}
\caption{
Difference between the dipole expectation values obtained with the \hbox{RMT-1st} and \hbox{RMT-6st} models relative to
the results obtained with \hbox{RMT-3st}  as a function of time for the laser parameters used to produce Fig.~\ref{fig:fig4} (central wavelength 1,064$\,$nm with a sin$^2$ envelope of the electric field and peak intensity of $\rm 5 \times 10^{13}\,W/cm^2$).
}
\label{fig:fig4dipolediff}       
\end{figure}

To further illustrate the numerical challenges associated with an accurate calculation of the spectrum, 
we show in Fig.~\ref{fig:fig4dipole} the induced dipole moment for this case and in Fig.~\ref{fig:fig4dipolediff} the
differences in the models relative to the \hbox{RMT-1st} result.  The situation is similar, but even more pronounced, 
to what we saw for the 248.6~nm case: a tiny difference in the calculated dipole moment can change the predicted 
height of the plateau by several orders of magnitude.  
 
\begin{figure}[t!]
\resizebox{0.480\textwidth}{!}{\includegraphics{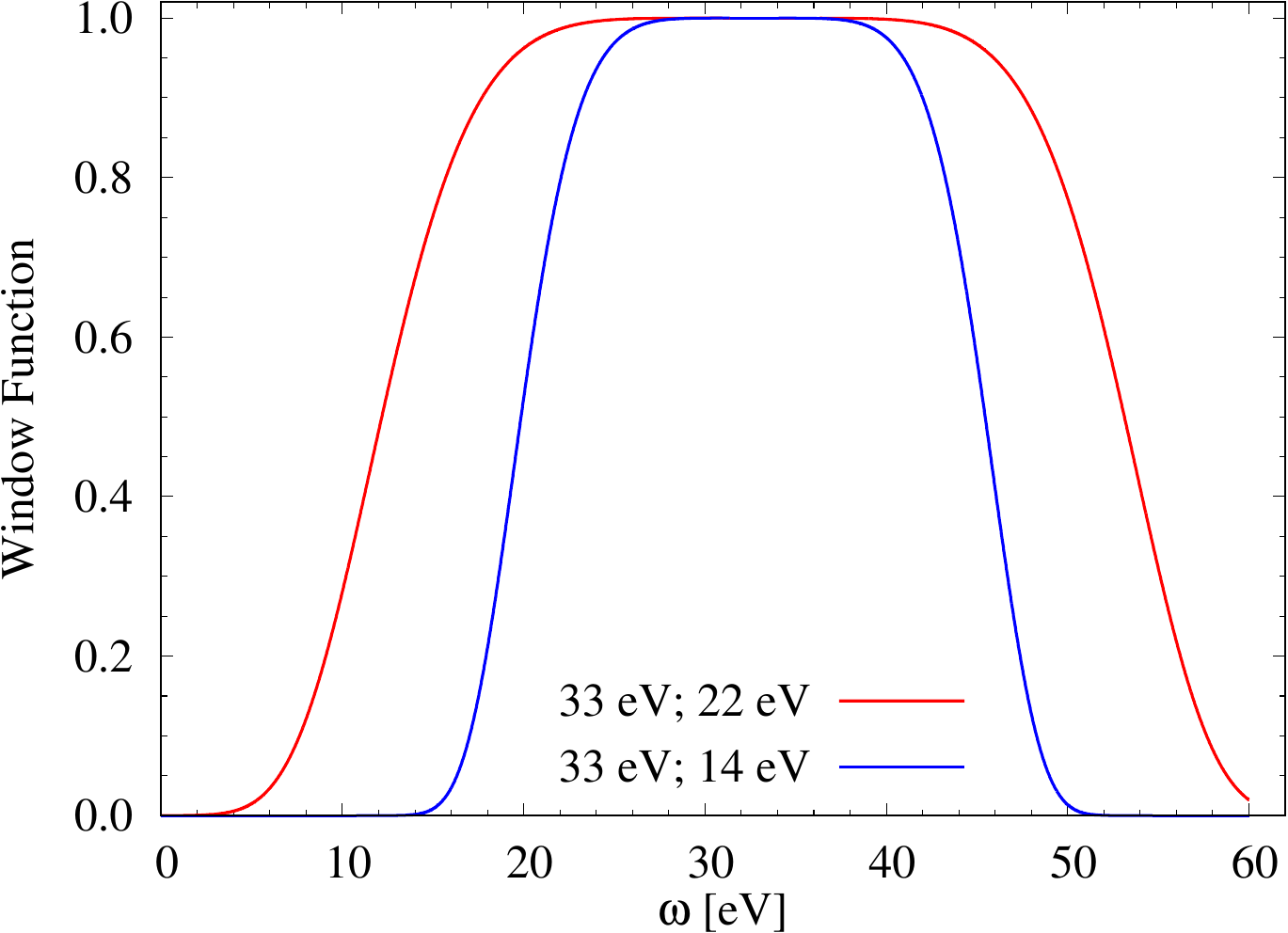}
}
\caption{
Super-Gauss window function centered on the plateau with different widths to focus on the frequencies of the plateau in Fig.~\ref{fig:fig4}. 
}
\label{fig:window}       
\end{figure}

Having seen that the differences in the dipole moment, as small as they are,
are still large enough to result in significantly different HHG spectra, we decided to investigate how much of the induced dipole moment with the laser parameters of Fig.~\ref{fig:fig4} is actually associated with the plateau of the spectrum.  We did this by applying a Super-Gauss window function 
\begin{equation}
W(\omega) \equiv {\rm e}^{-\{(\omega-\omega_{\rm c})/\omega_{\rm w}\}^6},     
\end{equation}
where $\omega_{\rm c}$ is the central frequency and $\omega_{\rm w}$ is the width, to the Fourier transform of the dipole moment and then calculating the
inverse transform. We centered the function around 33~eV and chose widths of 22~eV and 14~eV, respectively.  These functions are depicted in Fig.~\ref{fig:window}.
We chose a linear scale on purpose to demonstrate how an apparent ``zero'' can be misleading.

\begin{figure}[t!]
\resizebox{0.480\textwidth}{!}{\includegraphics{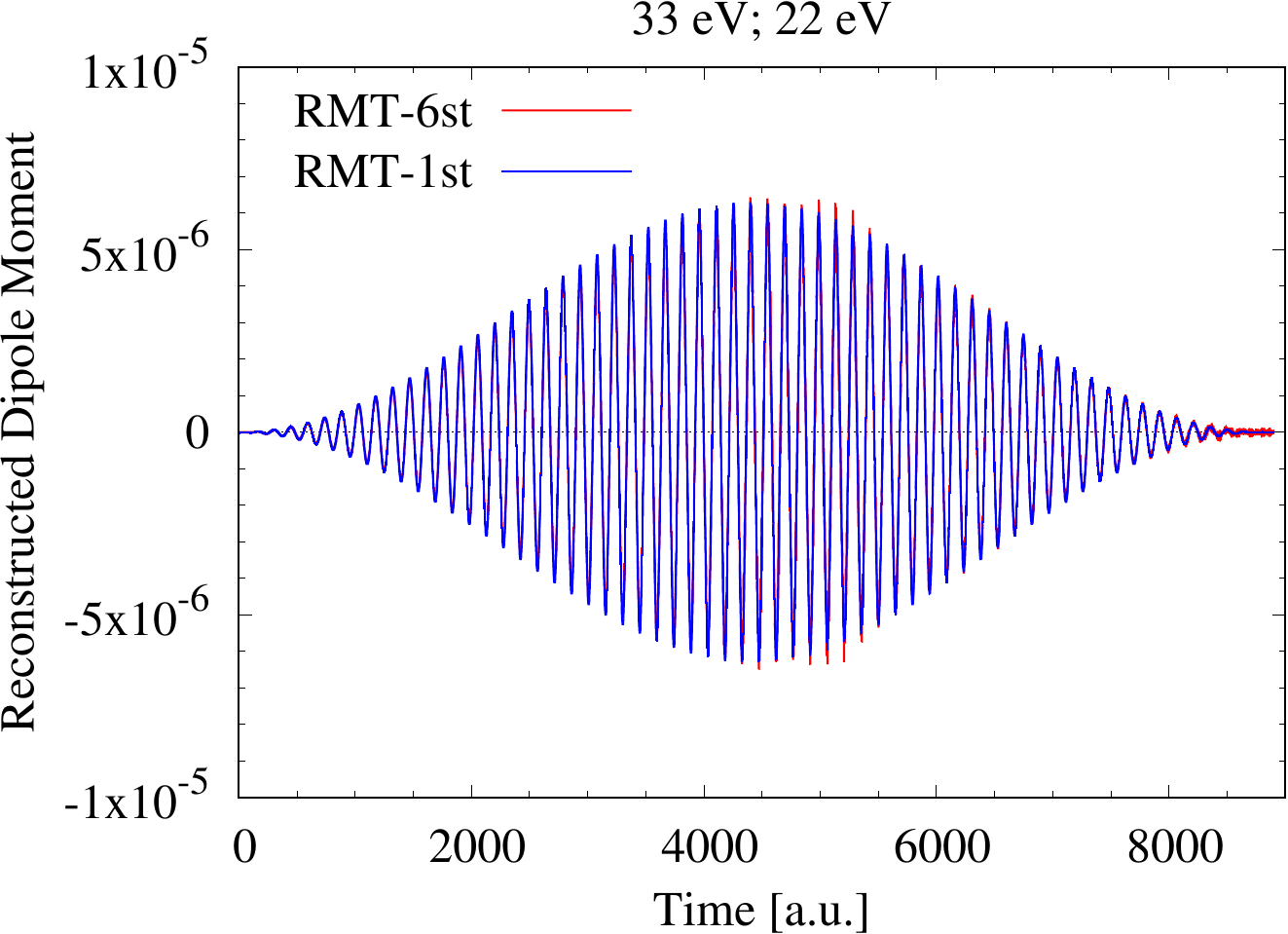}
}
\caption{
Reconstructed part of the dipole moment with the wider Super-Gauss window function centered on the plateau of Fig.~\ref{fig:fig4}. 
}
\label{fig:dipole-window-wide}       
\end{figure}

Figure~\ref{fig:dipole-window-wide} shows the reconstructed dipole moment with the window function. Using the wider function, the dipole moment is 
reduced by four orders of magnitude, but it is still dominated by the fundamental frequency. Consequently, even though the contribution of the fundamental is drastically (but still insufficiently) reduced in the HHG spectrum, the \hbox{RMT-1st} and RMT-6st models still reproduce the partial dipole moment in almost the same way. 

\begin{figure}[t!]
\resizebox{0.480\textwidth}{!}{\includegraphics{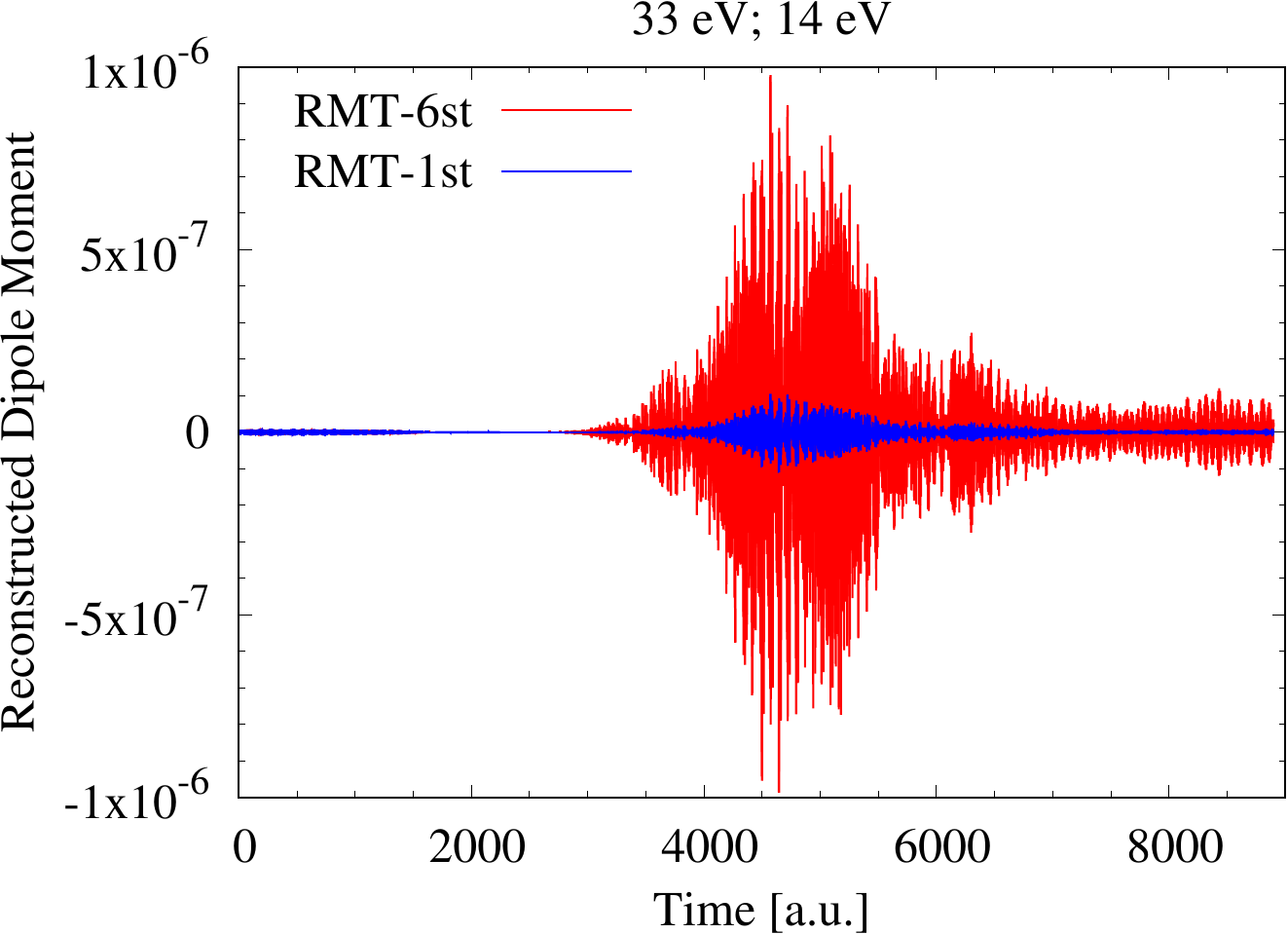}
}
\caption{
Reconstructed part of the dipole moment with the narrower Super-Gauss window function centered on the plateau of Fig.~\ref{fig:fig4}. 
}
\label{fig:dipole-window-narrow}       
\end{figure}

Figure~\ref{fig:dipole-window-narrow} shows the reconstructed dipole moment with the narrower function. Now the dipole moment is reduced by five orders of magnitude, the fundamental as well as the low-order harmonics are filtered out, and we see indeed a large difference in the reconstructed part of the dipole moment in the RMT-1st and RMT-6st models.  The \hbox{RMT-3st} reconstruction (not shown for clarity) is very similar to that obtained from RMT-1st.  While we cannot unambiguously decide that the RMT-6st results are problematic, these figures suggest that the problem is numerically ill-conditioned for this choice of parameters. For this very reason, however, we suggest it as a challenge for benchmark calculations to thoroughly test the many computer codes that have been used to predict {\it relative\/} rather than {\it absolute\/} HHG spectra. 

\begin{figure}[t!]
\resizebox{0.480\textwidth}{!}{\includegraphics{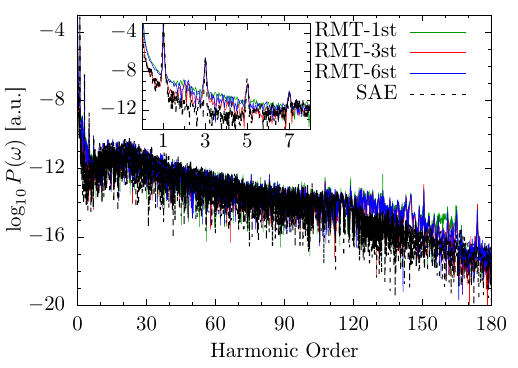}
}
\caption{
Same as Fig.~\ref{fig:fig4} for a peak intensity of $\rm 4 \times 10^{14}\,W/cm^2$.
The predicted cut-off energy from the three-step model is \hbox{$\approx\,160\,$eV}, corresponding to the 137$^{\rm th}\,$harmonic.
}
\label{fig:fig5}       
\end{figure}

\begin{figure}[t!]
\resizebox{0.480\textwidth}{!}{\includegraphics{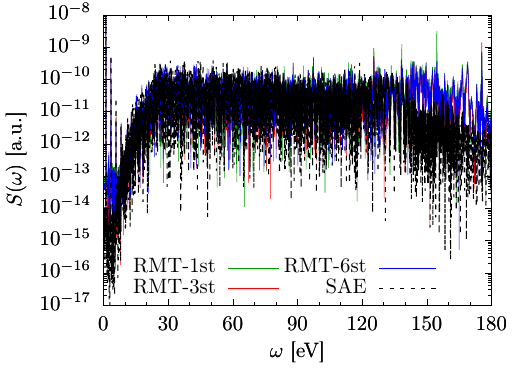}
}
\caption{
Spectral density as defined in Refs.~\cite{Joachain,PhysRevA.87.053406} for the laser parameters used to produce Fig.~\ref{fig:fig5}. The $y$-axis is logarithmic.
}
\label{fig:fig6}       
\end{figure}

Our final comparison with the results of Tong and Chu~\cite{Tong2001} is shown in Fig.~\ref{fig:fig5}, and we 
again also exhibit the results obtained with Eq.~(\ref{eq:joachain}) in Fig.~\ref{fig:fig6}. Even though details are 
hardly visible, there appears to be quantitative agreement with the results shown in Fig.~5(b)
of Ref.~\cite{Tong2001}.  

\begin{table*}[t!]
\begin{center}
\caption{Fraction (in \%) of the integral under various peaks in Fig.~\ref{fig:fig6} for 
a 60-cycle pulse of central wavelength 1,064$\,$nm with a sin$^2$ envelope 
of the electric field and peak intensities of $\rm Low = 5 \times 10^{13}\,W/cm^2$ and $\rm High = 4 \times 10^{14}\,W/cm^2$, respectively.
}
\vspace{0.2truemm}
\label{tab:tab1}
\resizebox{.79\textwidth}{!}{
\begin{tabular}{|l|c|c|c|c|c|c|c|c|}
\hline
Model & \multicolumn{2}{|c|}{SAE}  & \multicolumn{2}{|c|}{RMT-1st}  & \multicolumn{2}{|c|}{RMT-3st} & \multicolumn{2}{|c|}{RMT-6st}\Tstrut\\ 
\hline
Intensity      & Low &  High 
      & Low &  High 
      & Low &  High 
      & Low &  High  \Tstrut\\
\hline
fundamental           & 99.97689              &  9.71731   & 99.97233              &  7.61130 & 99.96863              &  9.35664                 & 99.96643              &  7.81866 \Tstrut\\
$3^{\rm rd}$ harmonic &  0.02309              &  0.42943   &  0.02764              &  0.25380 &  0.03134              &  0.33415                 &  0.03134              &  0.29346 \\
$5^{\rm th}$ harmonic &  0.00001              &  0.02763   &  0.00001              &  0.00681 &  0.00001              &  0.00859                 &  0.00001              &  0.00607 \\
plateau               &  $2.7\times 10^{-6}$  & 87.46749   &  0.00001              & 90.50232 &  0.00001              & 88.75972                 &  0.00014              & 89.97512 \\
\hline
\end{tabular}
}
\end{center}
\end{table*}

As pointed out in Ref.~\cite{Finger2022}, a quantity of significant practical interest is the conversion efficiency, i.e.,
what portion of the incident intensity can actually be converted into a few low-order harmonics and the plateau. 
In fact, a straight\-forward indication is a comparison of the integrals under the various peaks and the plateau.  Due to the 
likely singularity for $\omega \to 0$ in the definition of \hbox{Refs.~\cite{Tong2001,Guan2006}}, 
we only use Eq.~(\ref{eq:joachain}) to obtain the fractions relative to the integral under the entire curve. 

The results are listed in Table~\ref{tab:tab1}.  We only present them for the 1,064~nm case discussed in this subsection, 
since there is no real plateau for 248.6~nm.  For the relatively low peak intensity of $\rm 5 \times 10^{13}\,W/cm^2$, 
the spectra are completely dominated by the contribution from the
fundamental frequency, followed by rapid drops for the low-order harmonics that are still distinguishable on the graphs.
The relative importance of the plateau is most important according to the \hbox{RMT-6st} model
for this low intensity.  
However, the numbers are extremely small, and such small numbers are usually very difficult to calculate accurately.  

For the higher peak intensity of $\rm 4 \times 10^{14}\,W/cm^2$, on the other hand, the integral under the plateau
provides the dominant contribution to the spectrum. As one might expect for this case of larger numbers, 
the numerical challenges appear to be less significant. 
Table~\ref{tab:tab1} shows that for the high-intensity results, the plateau integrals all agree to within about 3\%, a significant improvement upon the order-of-magnitude disagreement observed in the low-intensity case. 
We note that the total ionization probability for the laser parameters
selected for this case is still only about 0.1\%, i.e., this is consistent with the usual HHG condition of almost negligible
ionization.

\section{Conclusions}
\label{sec:conclusions}
We have reported a comparison study for high-order harmonic generation in helium.  
While many results for this problem have been reported previously, we emphasized the importance of a number of
items that seem to have been frequently ignored in other works.  Most importantly, we continue to advocate that
theorists publish {\it absolute\/} numbers, together with the definition they used to obtain them. Furthermore, we illustrated that the induced dipole moment needs to be calculated with 
very high accuracy if reliable benchmark results are to be obtained, which can then be used to
assess the likely reliability of the predictions. 

Finally, we found a surprising effect on the
results for $\rm 5 \times 10^{13}\,W/cm^2$ by adding the \hbox{$n=3$} states (specifically the $3p$) of He$^+$ into the RMT model.  
We hope that the present work will encourage other
groups to perform calculations that, hopefully, will lead to established benchmark results. 
Since the underlying quantity is the induced dipole moment, from which all the other parameters (time derivatives, Fourier transforms, spectra in various conventions) can be calculated, we uploaded our results to \href{https://github.com/Abaw1397/He-HHG}{https://github.com/Abaw1397/He-HHG}.  To simplify the file structure, all the dipole moments were interpolated to a common time grid with a step size of 0.1 and are given from the beginning to the end of the driving pulse.  
We checked that the spectra shown in this paper can be reproduced with sufficient accuracy from these interpolated results.   

\section*{Acknowledgements} 
The work of A.T.B., S.S., J.C.d.V, K.R.H., and K.B.\ was supported
by the NSF through Grant No.~PHY-2110023, 
by the XSEDE supercomputer allocation No.~PHY-090031, 
and by the Frontera Pathways Project No.~PHY-20028.
A.T.B.\ is grateful for funding through NSERC via a Michael Smith Scholarship
to visit Drake University.
The work of N.D.\ was supported by the NSF through Grant No.~PHY-2012078.

 \bibliographystyle{apsrev4-1}

\end{document}